\documentclass[11pt]{article}
\input{epsf}
\usepackage{subfigure}
\usepackage{multirow}

\usepackage[utf8]{inputenc}
\usepackage{times}

\begin{document}
\addtolength{\baselineskip}{0.005in}

\vspace*{0.5cm}

\begin{center}
{\large\bf A Novel Analytic Approach to Model Line Edge Roughness using Stochastic Exposure Distribution in Electron-beam Lithography}
\end{center}

\vspace{0.1cm}
\begin{center}
Rui Guo and Soo-Young Lee \\
Department of Electrical and Computer Engineering \\
Auburn University, Auburn, AL, USA \\
\end{center}

\bigskip

\begin{center}
{\bf Abstract}
\end{center}
The line edge roughness (LER) becomes a issue of e-beam lithography when feature size is reduced into nanometers. Therefore, minimizing the LER is a important method to increase the density of circuit patterns. One of the possible ways is through simulation. The stochastic exposure distributions in the resist is generated by the Monte Carlo simulation. In addition a resist development simulation needs to be carried out. Although there are several ways to simulate or estimate LER but none of them can reveal as much the inner relationship between LER and different parameters as theocratical analysing methods can do. In this paper, a new approach to analytically derive the LER based on the statistical exposure, is described. Our approach is based on analytic model of stochastic exposure distribution, deriving standard deviation of exposure and analyzing the variance of edge location after development. Even though it may not be a complete modeling of LER, it can still show some strong relationship between LER and some inner parameters.

\vspace*{0.5cm}

{\em Keywords:} electron-beam lithography, line edge roughness, stochastic exposure distribution, point spread function

\addtolength{\baselineskip}{0.16in}

\section{Introduction}

Electron-beam (e-beam) lithography is widely employed in a variety of areas such as fabrication of photomasks, imprint lithography molds, experimental circuit patterns, etc., for its great ability to transfer ultra-fine features onto the resist and eventually to the substrate material. Its main limitations are the low outputs due to the pixel-by-pixel or feature-by-feature writing and the proximity effect caused by electron scattering. The importance of developing effective and efficient schemes for correcting the proximity effect has been well recognized for a long time, and various methods were proposed and implemented by many researchers \cite{Osawa2001}-\cite{Zhao2014}. As the feature size decreases well below microns into nanoscale (100 nm or less), the line edge roughness (LER) can significantly affect the minimum feature size and the maximum circuit density realizable in practice. One of the reasons is the LER in a pattern transferred onto the resist by electron-beam (e-beam) lithographic process is independent of the feature size. Therefore, it is unavoidable to minimize the LER in order to maximize the feature density in a pattern of nanoscale features. One important step required in developing an effective method to minimize the LER is to analyze and estimate the LER. A possible approach to such analysis is to rely on the Monte Carlo simulations \cite{Zhou2006}-\cite{Demers2011}. While simulation approach is flexible, the main drawback is that it is computationally intensive, in particular the step of simulating resist development.

In this paper, a new analytic approach based on e-beam lithography, is generated to estimate the LER in order to avoid the repetitive time-consuming simulation. Based on the moments of PSF's such as mean and variance, the stochastic fluctuation of exposure (energy deposited) in the resist is derived. The objective of this study is to develop an analytic method for estimating the LER caused by the exposure fluctuation only. It should be pointed out that the method is still applicable even when other factors are to be also generated by others.

The rest of the paper is organized as follows. The analytic model and the definition of terms and the assumptions of the system are described in Sec. \ref{sec:Model}. Deriving LER from the moments of PSF is illustrated in Sec. \ref{sec:LER}. Results are discussed in Sec. \ref{sec:Results}. A summary is provided in Sec. \ref{sec:Sum}.

\section{Analytic Model}
\label{sec:Model}

In our model, the stochastic distribution of exposure is derived first. In addition the distribution of developing rate which is critical to trace the paths and derive the variance (LER) of boundaries of resist profile after development, is derived from the distribution of exposure. Finally by using the concept of developing path which models the resist development process \cite{Dai2014}, the LER is calculated and a scale factor is derived to make the result realistic.
\begin{figure}[htb]
 \centering
 \epsfxsize=8cm
\ \epsfbox{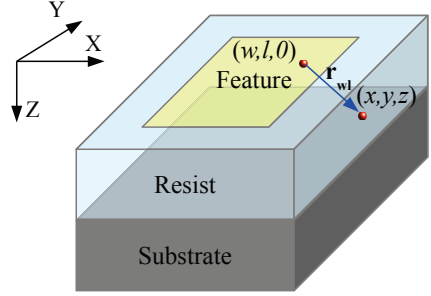} \\
\vspace{-0.2cm}
 \caption{A 3-D model of substrate system consists of resists, substrate and a long line feature.
}
\label{fig:Model}
\end{figure}

\begin{figure}[htb]
 \centering
 \epsfxsize=7cm
\ \epsfbox{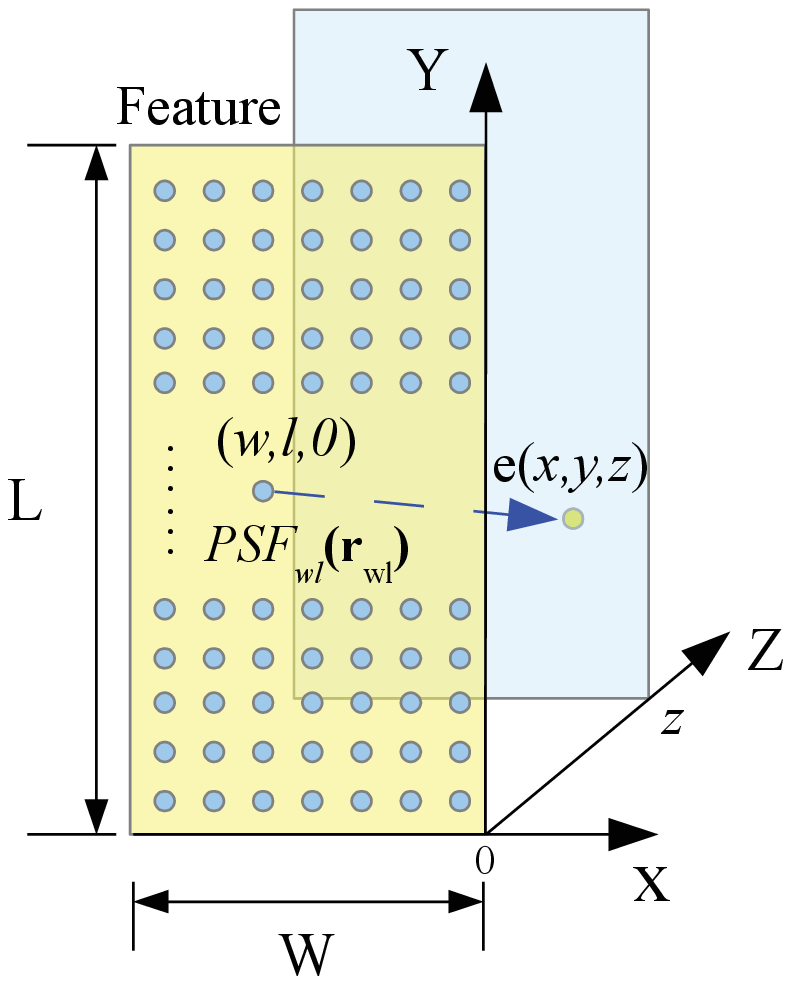} \\
\vspace{-0.2cm}
 \caption{Long line feature with the size of $L\times W$, and the exposure of the points needs to be calculated.
}
\label{fig:Exposure}
\end{figure}

\begin{figure}[!htb]
 \centering
 \epsfxsize=5cm
\ \epsfbox{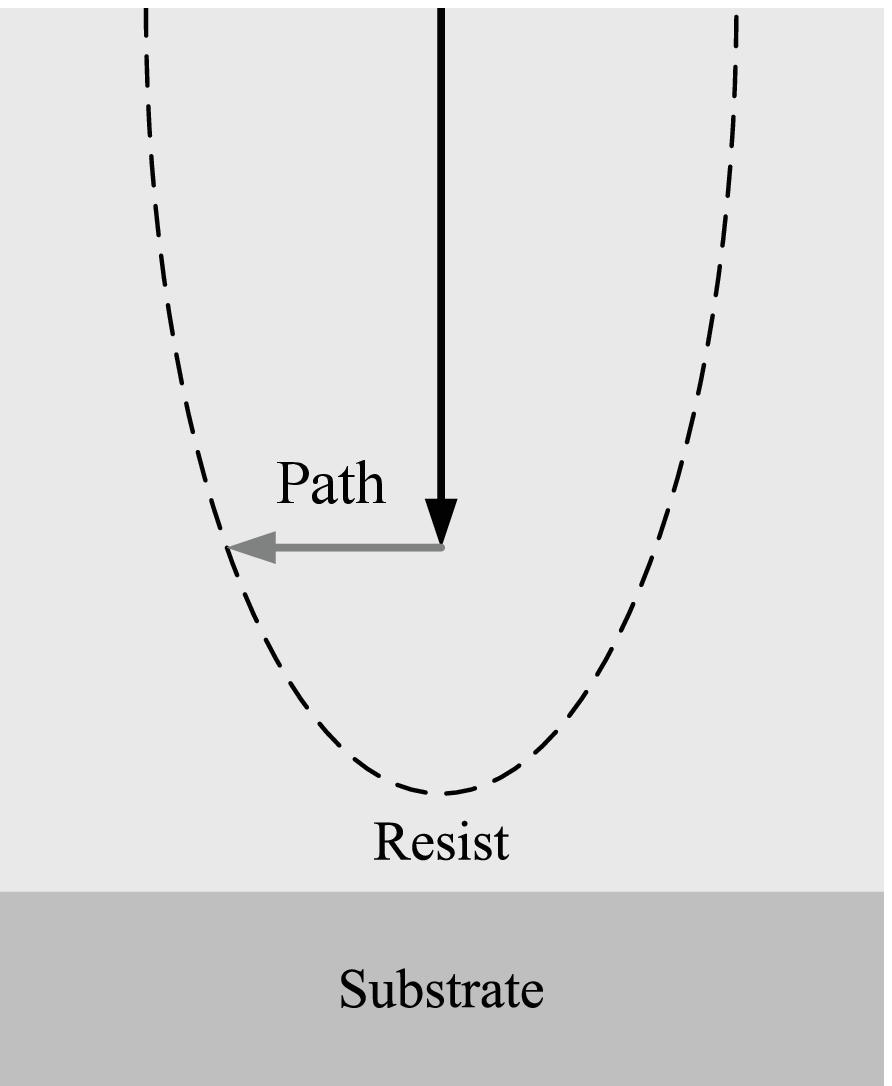} \\
(a)\\
 \epsfxsize=5cm
\ \epsfbox{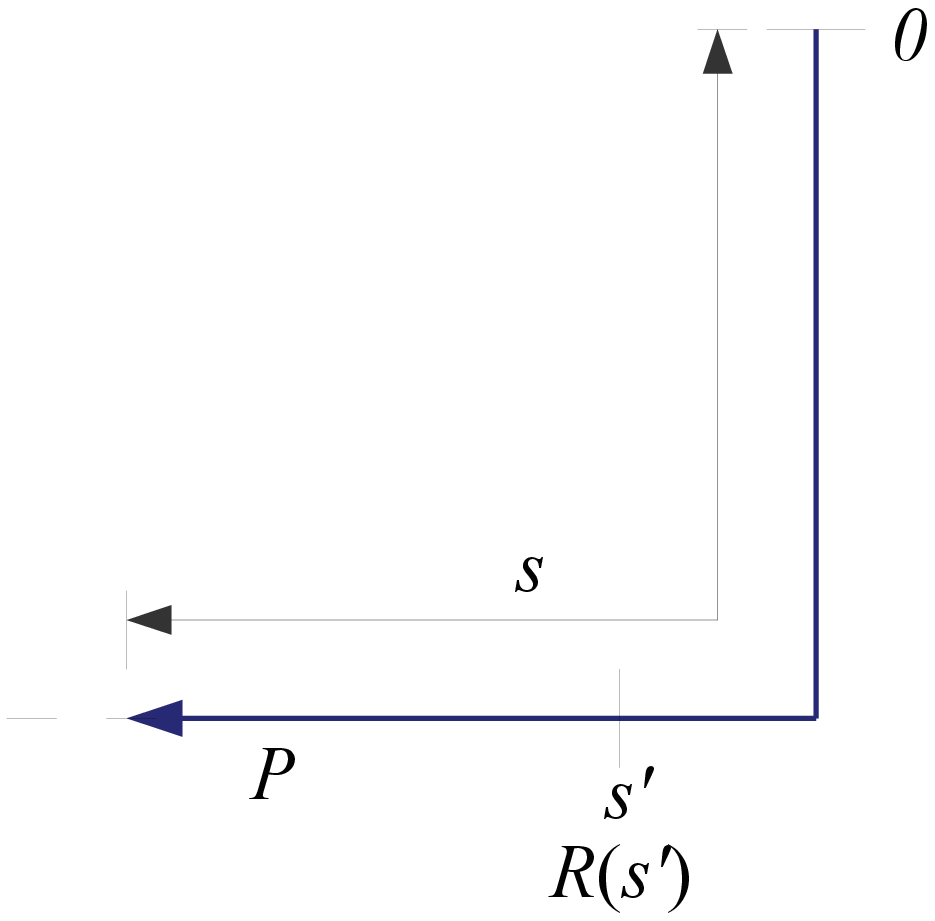} \\
(b)\\
\vspace{-0.2cm}
 \caption{Description of path concept in the resist: (a), and detail division on the path (b).
}
\label{fig:Path}
\end{figure}

A typical substrate system which consists of a substrate and a certain type of resist with initial thickness on top of the substrate, is employed in this study as illustrated in Fig. \ref{fig:Model}, where the resist depth is along the $z$-dimension. The feature is $W$ pixels wide in $x$ dimension and $L$ pixels long in the $y$ dimension and its right edge locates on the $y$ axis, as shown in Fig. \ref{fig:Exposure}. Let $PSF_{wl}(x,y,z)$ denote the point spread function (PSF) which depicts the exposure distribution when point $(w,l,0)$ is exposed \cite{Seo2000}-\cite{Unal2011}, where $w$ is from $-1$ to $-W$ and $l$ is from $1$ to $L$. Let $e(x,y,z)$ denote exposure and $R(x,y,z)$ denote the resist developing rate at point $(x,y,z)$. According to the certain relationship between exposure and resist developing rate, i.e. conversion formula, the $R(x,y,z)$ at any point can be derived from that point's $e(x,y,z)$. For any path in the resist as shown in Fig. \ref{fig:Path}, where $R_1,R_2,\cdots$, denote the resist developing rates following the path and $u$ and $v$ are the lateral length and vertical length of the path, the mean and the standard deviation of edge location for each path can be derived from $1/R_1,1/R_2,\cdots$ given the developing time $T$. After scaling to compensate effects along $y$ axis, path with largest edge location will be selected as the dominant path and its variance can be considered as the LER.

In the e-beam lithography, writing is processed point by point and the processes of different points are independent of each other. So $PSF_{wl}(x,y,z)$ and $PSF_{w'l'}(x,y,z)$  are independent of each other when $w\neq w'$ or $l\neq l'$ for they are corresponding to the process of different points. For any point $(x,y,z)$, assume $PSF_{wl}(x,y,z)$ is Gaussian distributed with mean $m_{PSF}(x,y,z)$ and variance $\sigma_{PSF}^2(x,y,z)$. Assume each $PSF_{wl}(x,y,z)$ has the same mean $m_{PSF}(x,y,z)$ and variance $\sigma_{PSF}^2(x,y,z)$ under the same condition such as using the same dose, under the same shot noise distribution, using the same beam diameter. Terms which will be used in the following parts are redefined as follows.

\begin{itemize}
\item
$PSF_{wl}(x,y,z)$: point spread function which depicts the energy distribution when point $(w,l,0)$ is exposed
\item
$m_{PSF}(x,y,z)$: the mean of the set of PSF's at point $(x,y,z)$.
\item
$\sigma_{PSF}(x,y,z)$: the standard deviation of PSF's at point $(x,y,z)$.
\item
$e(x,y,z)$: exposure at the point $(x,y,z)$.
\item
$R(x,y,z)$: resist developing rate at the point $(x,y,z)$.
\item
$R_k$: resist developing rate on the $kth$ pixel on the path.
\item
$T$: developing time.
\item
$f()$: probability density function.
\end{itemize}

\section{Estimation of LER from PSF}
\label{sec:LER}

\subsection{From PSF to exposure fluctuation}

According to definition, for any given point $(x,y,z)$ in the resist, the exposure of that point can be expressed as $e(x,y,z)=\sum_w\sum_lPSF_{wl}(x-w,y-l,z)$. Since $(x,y,z)$ stay the same and only $w$ and $v$ vary from $-1$ to $-W$ and from $1$ to $L$, it would be easier to use a vector $\textbf{r}_{wl}$ which starts from Point $(w,l,0)$ to Point $(x,y,z)$ and $\textbf{r}_{wl}$ = $(x-w,y-l,z)$. Thus $e(x,y,z)=\sum_w\sum_lPSF_{wl}(\textbf{r}_{wl})$. Since it is a linear relationship between PSF's and exposure, mean of exposure can be directly derived as

\begin{equation}
    E\{e(x,y,z)\}=\sum_{w=-1}^{-W}\sum_{l=1}^LE\{PSF_{wl}(\textbf{r}_{wl})\}=\sum_{w=-1}^{-W}\sum_{l=1}^Lm_{PSF}(\textbf{r}_{wl})
\label{eqn:meanofe}
\end{equation}

\begin{eqnarray}
    \sigma^2(e(x,y,z))&=&E\{e^2(x,y,z)\}-[E\{e(x,y,z)\}]^2\nonumber\\
    &=&E\{\sum_{w'}\sum_{l'}\sum_w\sum_lPSF_{w'l'}(\textbf{r}_{w'l'})PSF_{wl}(\textbf{r}_{wl})\}-[\sum_w\sum_lm_{PSF}(\textbf{r}_{wl})]^2\nonumber
\end{eqnarray}

For $w\neq w'$ or $l\neq l'$, $PSF_{w'l'}(\textbf{r}_{w'l'})$ and $PSF_{wl}(\textbf{r}_{wl})$ are corresponding to different points in the feature and thus they are independent. Therefore the part $E\{\sum_{w'}\sum_{l'}\sum_w\sum_lPSF_{w'l'}(\textbf{r}_{w'l'})PSF_{wl}(\textbf{r}_{wl})\}$ can be expanded to $E\{{\sum_{w'}\sum_{l'}\sum_w\sum_l}_{~w\neq w'or l\neq l'}PSF_{w'l'}(\textbf{r}_{w'l'})PSF_{wl}(\textbf{r}_{wl})\}+E\{\sum_w\sum_l[PSF_{wl}(\textbf{r}_{wl})]^2\}$ which is equal to ${\sum_{w'}\sum_{l'}\sum_w\sum_l}_{~w\neq w'or l\neq l'}m_{PSF}(\textbf{r}_{w'l'})m_{PSF}(\textbf{r}_{wl})+\sum_w\sum_lE\{[PSF_{wl}(\textbf{r}_{wl})]^2\}$.
The part $[\sum_w\sum_lm_{PSF}(\textbf{r}_{wl})]^2$ can be expanded to a similar form ${\sum_{w'}\sum_{l'}\sum_w\sum_l}_{~w\neq w'or l\neq l'}m_{PSF}(\textbf{r}_{w'l'})m_{PSF}(\textbf{r}_{wl})+\sum_w\sum_lm_{PSF}^2(\textbf{r}_{wl})$. Eliminate the common part and the final result is

\begin{eqnarray}
    \sigma^2(e(x,y,z)) &=&\sum_{w=-1}^{-W}\sum_{l=1}^L[E\{[PSF_{wl}(\textbf{r}_{wl})]^2\}-m_{PSF}^2(\textbf{r}_{wl})]\nonumber\\
    &=&\sum_{w=-1}^{-W}\sum_{l=1}^L\sigma_{PSF}^2(\textbf{r}_{wl})
\label{eqn:varofe}
\end{eqnarray}

The result demonstrates that the variance of exposure at any point simply equals to the convolution of the variance of the set of PSF's with certain feature.

\subsection{From exposure fluctuation to rate fluctuation}

The developing rate $R(x,y,z)$ at each point is calculated from its exposure $e(x,y,z)$ through a nonlinear mapping of (e-to-R) conversion formula which is experimentally determined. The conversion formula derived in our experiment is given by:
\begin{equation}
    R(x,y,z) = C(e(x,y,z)) = 20000 exp\left[\left(\frac{e(x,y,z)-6.0e10}{1.8e10}\right)^2\right] - 0.29
\label{eqn:covert}
\end{equation}

Since the relationship (conversion formula) between exposure and developing rate is not linear, mean and variance of rate ($R(x,y,z)$) or one over rate which will be used in the next step can not be derived directly. But since the conversion formula is a monotone function, if the probability density function of exposure ($f_e$) can be derived, the probability density function of developing rate ($f_R$) can also be derived from theorem:

\begin{eqnarray}
    f_R(R(x,y,z)) = f_e(C^{-1}(R(x,y,z)))\left|\frac{d}{dR(x,y,z)}C^{-1}(R(x,y,z))\right|
\label{eqn:thoerem}
\end{eqnarray}

According to the central limit theorem, as exposure is calculated by convolution which equals to the sum of a large number of PSF's, exposure should be approximately treated as Gaussian distributed with mean of equation (\ref{eqn:meanofe}) and variance of equation (\ref{eqn:varofe}). Base on equation (\ref{eqn:covert}) and equation (\ref{eqn:thoerem}), the distribution of rate can be derived.

\subsection{From rate fluctuation to the LER}
\label{subsec:ler}
On each layer of resist, the longest path (with largest mean value) determine the edges of feature and its variation determines the LER. In terms of $R(x,y,z)$, the developing time $T$ can be expressed as a line integral as shown in fig. \ref{fig:Path}:

\[
    T = \int_0^{s_0}\frac{ds}{R(x,y,z)}
\]

Assume a Gaussian distribution of $1/R(x,y,z)$ with $m_{1/r}(x,y,z)$ and variance $\sigma_{1/R}^2(x,y,z)$ which can be calculated from the distribution of $R(x,y,z)$ and its independence between $(x,y,z)$ and $(x',y',z')$. Then $f(T|s)$ is also a Gaussian distribution.

\[
    m(T|s_0) = \int_0^{s_0}m_{1/R}(x,y,z)ds
\]

\[
    \sigma^2(T|s_0) = \int_0^{s_0}\sigma_{1/R}^2(x,y,z)ds
\]

Since the conversion formula is an exponential function, assuming mean and variance of exposure are linear to $s$ when close to edge of the feature, the $m_{1/R}$ and $\sigma_{1/R}^2$ should also be an exponential function of $s$. As $m(T|s_0)$ and $\sigma^2(T|s_0)$ are the integral of an exponential function, they can also be expressed as an exponential form as
\[
    m(T|s_0) = a_1e^{b_1s_0}
\]

\[
    \sigma^2(T|s_0) = a_2e^{b_2s_0}
\]

The $a_1$, $b_1$, $a_2$, and $b_2$ are the constants dependent on the feature pattern and PSF shapers. For any $s$, the distribution of develop time given edge location $f(T|s)$ can be derived. Instead of the distribution of develop time given edge location, the distribution of edge location given develop time $T$ ($f(s|T)$) is eventually needed to derive the mean edge location and variance of edge location given time $T$. Based on the  mean edge location given time $T$, the longest path can be found and its standard deviation of edge location given $T$ can be considered as LER. To derive $f(s|T)$, according to Bayes' Theorem,

\begin{eqnarray}
    f(s|T)=\frac{f(T|s)f(s)}{f(T)}
\label{eqn:bayes}
\end{eqnarray}

The distribution of developing time $f(T)$ can be assumed or tested through experiment, i.e., $T$ is a Gaussian distribution with mean $\mu_T$ and variance $\sigma^2_T$. The distribution of edge location $f(s)$ cannot be directly assumed or tested but can be derive from $f(T)$ if the mapping from $T$ to $s$ can be derived. Here the relationship between $T$ and $s$ is approximated by the curve $m(T|s)$ which describes mean develop time given each location. Meanwhile $m(T|s)$ can also be interpreted as given develop time what is the mean edge location should be. Since $m(T|s)$ is a monotonous function, the probability density function of edge location $f(s)$ can be derived as:

\begin{eqnarray}
    f_s(s)= f_T(m(T|s))\frac{dm(T|s)}{ds} \nonumber
\end{eqnarray}

\begin{figure}[!htb]
 \centering
 \epsfxsize=7cm
\ \epsfbox{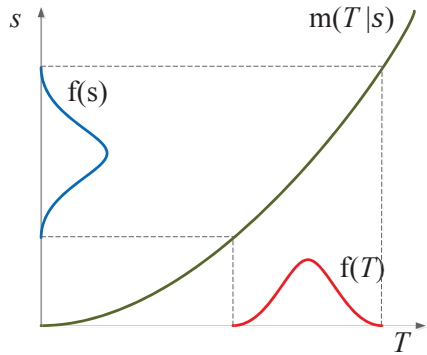} \\
\vspace{-0.2cm}
 \caption{illustration of deriving distribution of $s$ from distribution of $T$ and $m(T|s)$
}
\label{fig:dist_s}
\end{figure}

This process is illustrated in the Fig. \ref{fig:dist_s}. Put $f(T)$, $f(s)$, and $f(T|s)$ into equation (\ref{eqn:bayes}), $f(s|T)$ can be derived and the mean edge location and variance of edge location given developing time can be derived as:

\[
    E\{s|T\}=\int_0^\infty sf(s|T)ds
\]

\[
    Var\{s|T\}=\int_0^\infty s^2f(s|T)ds - E^2\{s|T\}
\]

After simplification, use denotations $A_T$ and $B_T$ as:

\[
    A_T=\frac{a_1b_1}{\sqrt{2\pi a_2}}e^{\frac{T^2-2\mu_T}{2\sigma_T^2}}e^{-\frac{T^2}{2a_2}-\frac{a_1^2}{2a_2}+ \frac{2Ta_1}{2a_2}-\frac{a_1}{2\sigma_T^2}+\frac{\mu_Ta_1}{\sigma_T^2}}
\]

\[
    B_T= -\frac{b_2T^2}{2a_2} +\frac{(2b_1-b_2)a_1^2}{2a_2} -\frac{(b_1-b_2)a_1T}{2a_2} +\frac{a_1b_1}{\sigma_T^2} -\frac{\mu_Ta_1b_1}{\sigma_T^2}-(b_1-\frac{b_2}{2})
\]

The final result can be expressed as:

\[
    E\{s|T\}=\frac{A_T}{B_T}
\]

\[
    Var\{s|T\}=\frac{2A_T-A_T^2}{B_T^2}
\]

\subsection{Adjustment}

The calculation in section \ref{subsec:ler} only includes the effects along $x$ dimension and its result is larger than the real LER. One reason is the effects along $y$ dimension (adjacent path) will compensate part of the difference and make the LER relatively smaller. To make this compensation, $\rho_T$ which is the correlation of develop time along adjacent path is used as an indicator. $\rho_T$ equals to -1 means two adjacent paths are always having different developing Time at same location and should make more compensation. Otherwise $\rho_T$ equals to 1 means two adjacent paths are always having same developing Time and thus no compensation should be made at all. When  $\rho_T$ equals to 0, it is total random and two cases can happen at equal chance.

\begin{figure}[!htb]
 \centering
 \epsfxsize=7cm
\ \epsfbox{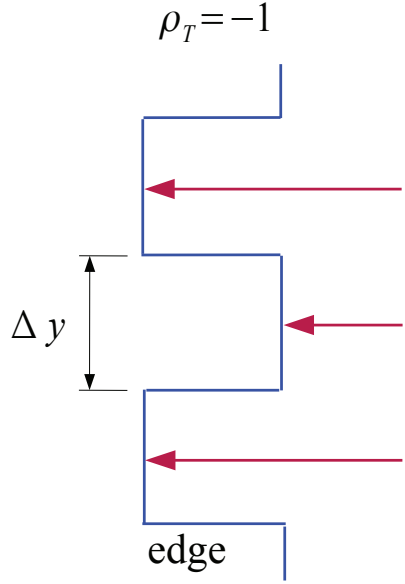} \\
\vspace{-0.2cm}
 \caption{ Demonstration of effects along $y$ dimension when correlation of developing time between adjacent path equals to -1.
}
\label{fig:scale}
\end{figure}

When $\rho_T$ equals to -1, as shown in Fig. \ref{fig:scale}, difference of developing time ($\Delta T$) which is equal to $\sqrt{2}\sigma(T|x)$ and average developing rate ($R_k$) needs to be used. When $\Delta T\times R_k$ is larger than the one pixel size ($D_x$), all difference will be compensate. If it is smaller than one pixel size, then $1-({2R\Delta T}/{D_x})^2$ will be compensated to LER. So the final scale factor is the equation:

\begin{eqnarray}
     \frac{1+\rho_T}{2}&&      ~~~~~~~~~~~~                   when~~    \frac{2R\Delta T}{D_x} \geq 1 \nonumber\\
     \left(1-\left(\frac{2R\Delta T}{D_x}\right)^2\right)\cdot\frac{1-\rho_T}{2} &+& \frac{1+\rho_T}{2}  ~~when~~    \frac{2R\Delta T}{D_x} < 1
\label{eqn:scale}
\end{eqnarray}

${(1-\rho_T)}/{2}$ and ${(1+\rho_T)}/{2}$ indicate the chances that two cases can happen. To calculate $\rho_T$, $Cov(T)$ and $\sigma^2(T)$ need to be calculated first. $Cov(T)$ and $\sigma^2(T)$ can be expressed as sums of $Cov(1/R_k)$ and sums of $\sigma^2(1/R_k)$ as
$\sigma^2(T)=D_z^2\sigma^{2}_1+\cdots+D_z^2\sigma^{2}_{Z_p}+D_x^2\sigma^{2}_{Z_p+1}+\cdots+(s-kD_x)^2\sigma^{2}_{Z_p+k}$ and $Cov(T)=D_z^2Cov(1/R_1)+\cdots+D_z^2Cov(1/R_{Z_p})+D_x^2Cov(1/R_{Z_p+1})+\cdots+(s-kD_x)^2Cov(1/R_{Z_p+k})$. $Cov(1/R)$ can be derived from $Cov(e)$. Similar to equation (\ref{eqn:varofe}),

\begin{eqnarray}
    Cov(e(x,y,z))=\sum_w\sum_l Cov_{PSF}(\textbf{r}_{wl})\nonumber
\end{eqnarray}

Because the conversion formula is not linear, a piecewise linear method is used to approximate the conversion formula in each segment. The derivative in the middle point of each segment is used to approximate the slope and $Cov(1/R)$ = ${g'(e)}^2Cov(e)$, where $g(e) = 1/R$. The final LER can be expressed as:

$$ LER=\left\{
\begin{array}{rcl}
Var\{s|T\}\cdot\frac{1+\rho_T}{2} ~~~~~~~~~~~~~~~~~~~~~when~~    \frac{2R\Delta T}{D_x} \geq 1\\
Var\{s|T\}\cdot\left(\left(1-\left(\frac{2R\Delta T}{D_x}\right)^2\right)\cdot\frac{1-\rho_T}{2} + \frac{1+\rho_T}{2}\right) ~~when~~    \frac{2R\Delta T}{D_x} < 1
\end{array} \right. $$

\section{Results and Discussion}
\label{sec:Results}

\begin{figure}[!htb]
\label{fig:scaled}
 \centering
 \epsfxsize=10cm
\ \epsfbox{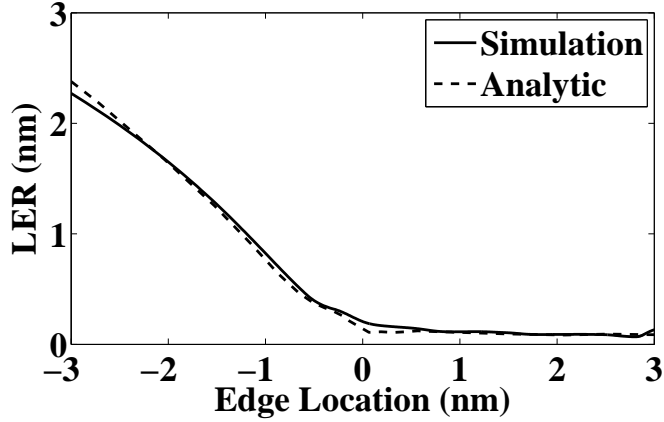} \\
\vspace{-0.2cm}
 \caption{scaled analytic results and simulation results at the bottom layer. Substrate system consists of 300nm PMMA on Si. The e-beam energy is 50keV and dose is 640 $\mu C/cm^2$, where pixel interval is 1nm and feature size is 0.08$\times$0.256 $\mu m^2$. PSF's are generated by 1K electrons.
}
\end{figure}

To verify the analytic method, LER derived from the exposure generated by Mont Carlo simulation and followed by resist development simulation is used for comparing. Resist is using 300nm PMMA on Si, the e-beam energy is 50kev and dose is 640 $\mu C/cm^2$ where pixel interval is 1 nm. The LER derived without scaling is much larger than the simulation results. It is because effects along $y$ dimension can significantly reduce the LER and the strength of the interaction depends on the difference of adjacent path. To appropriately adjust analytic results, correlation of developing time between adjacent paths is calculated and used as an index to show the difference between adjacent paths. Basically, larger difference would be required to compensate more and vice versa. Based on the scale factor described in equation (\ref{eqn:scale}), the results are shown in Fig. \ref{fig:scaled}. As can be seen in the figure, the results are much closer with each other.
\begin{figure}[!htb]
 \centering
 \epsfxsize=10cm
\ \epsfbox{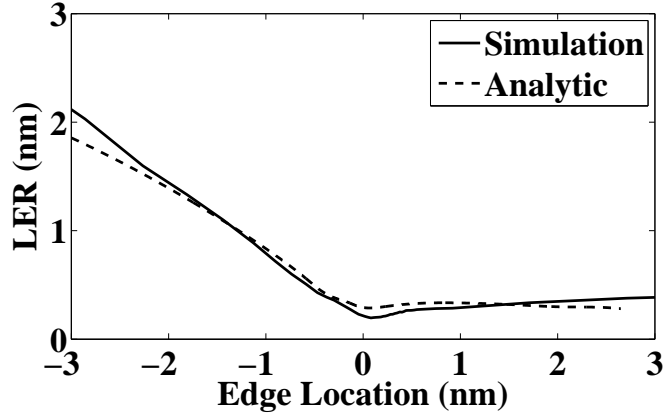} \\
\vspace{-0.2cm}
 \caption{LER estimated by the analytic model and simulation at the bottom layer. Substrate system consists of 300nm PMMA on Si. The e-beam energy is 50keV and dose is 640 $\mu C/cm^2$, where pixel interval is 1nm and feature size is 0.06$\times$0.256 $\mu m^2$. PSF's are generated by 40 electrons.}
\label{fig:40eLER}
\end{figure}

\begin{figure}[!htb]
 \centering
 \epsfxsize=10cm
\ \epsfbox{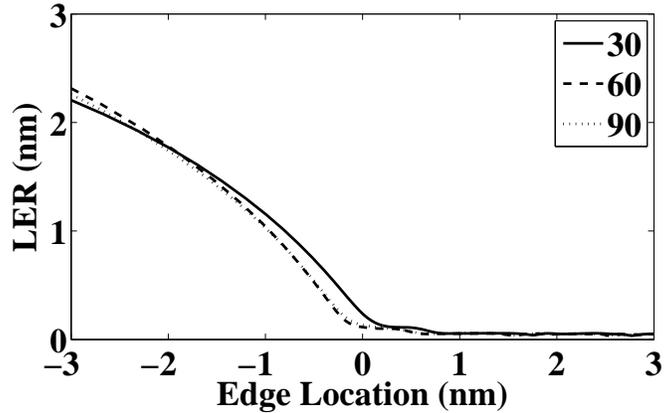} \\
\vspace{-0.2cm}
 \caption{LER results estimated by the analytic model with different feature size at the bottom layer. Substrate system consists of 300nm PMMA on Si. The e-beam energy is 50keV and dose is 640 $\mu C/cm^2$, where pixel interval is 1nm and feature sizes are 0.03$\times$0.256 $\mu m^2$, 0.06$\times$0.256 $\mu m^2$, 0.09$\times$0.256 $\mu m^2$ PSF's are generated by 1K electrons.
}
\label{fig:DiffSize}
\end{figure}

\begin{figure}[!htb]
 \centering
 \epsfxsize=10cm
\ \epsfbox{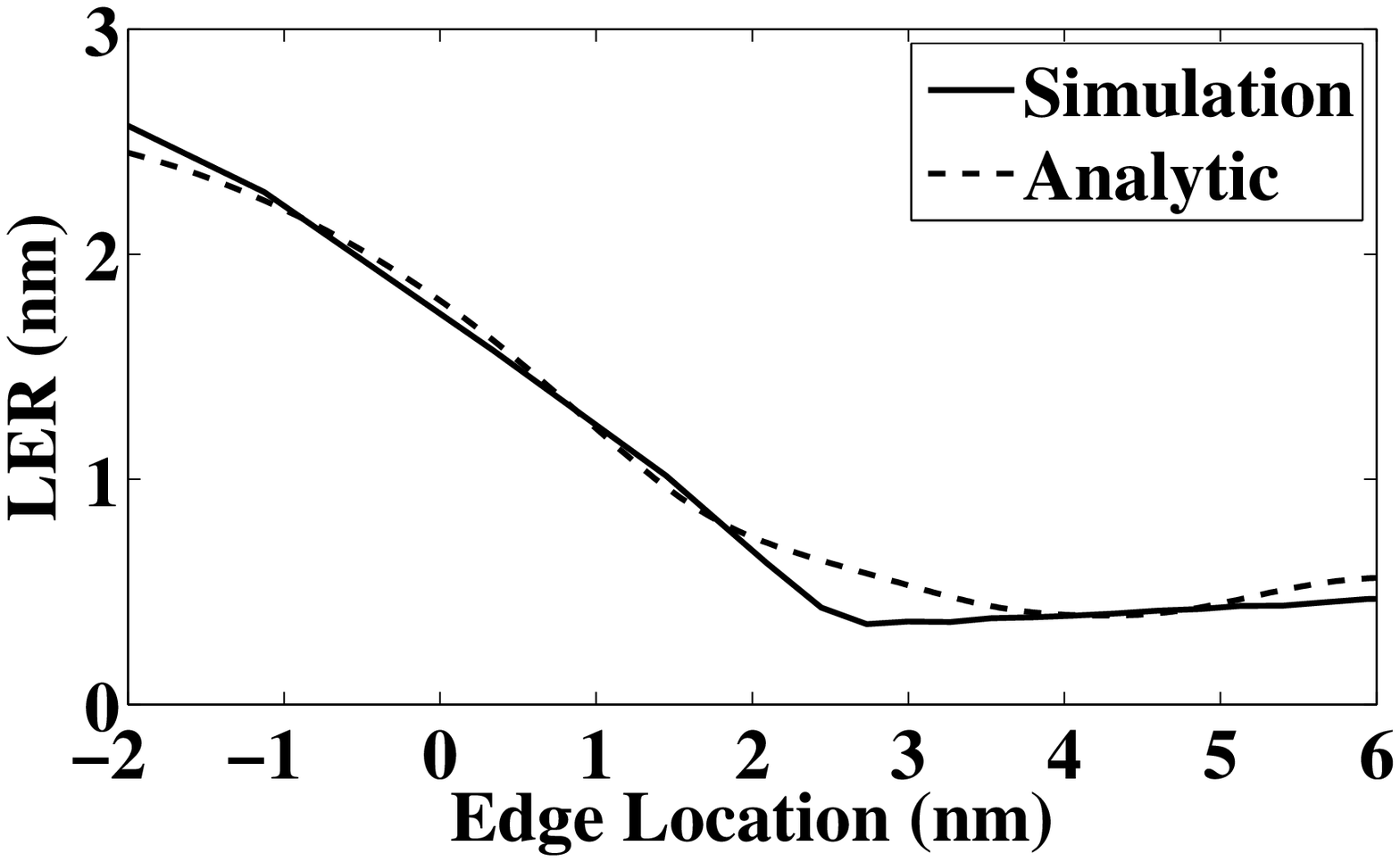} \\
(a)\\
 \epsfxsize=10cm
\ \epsfbox{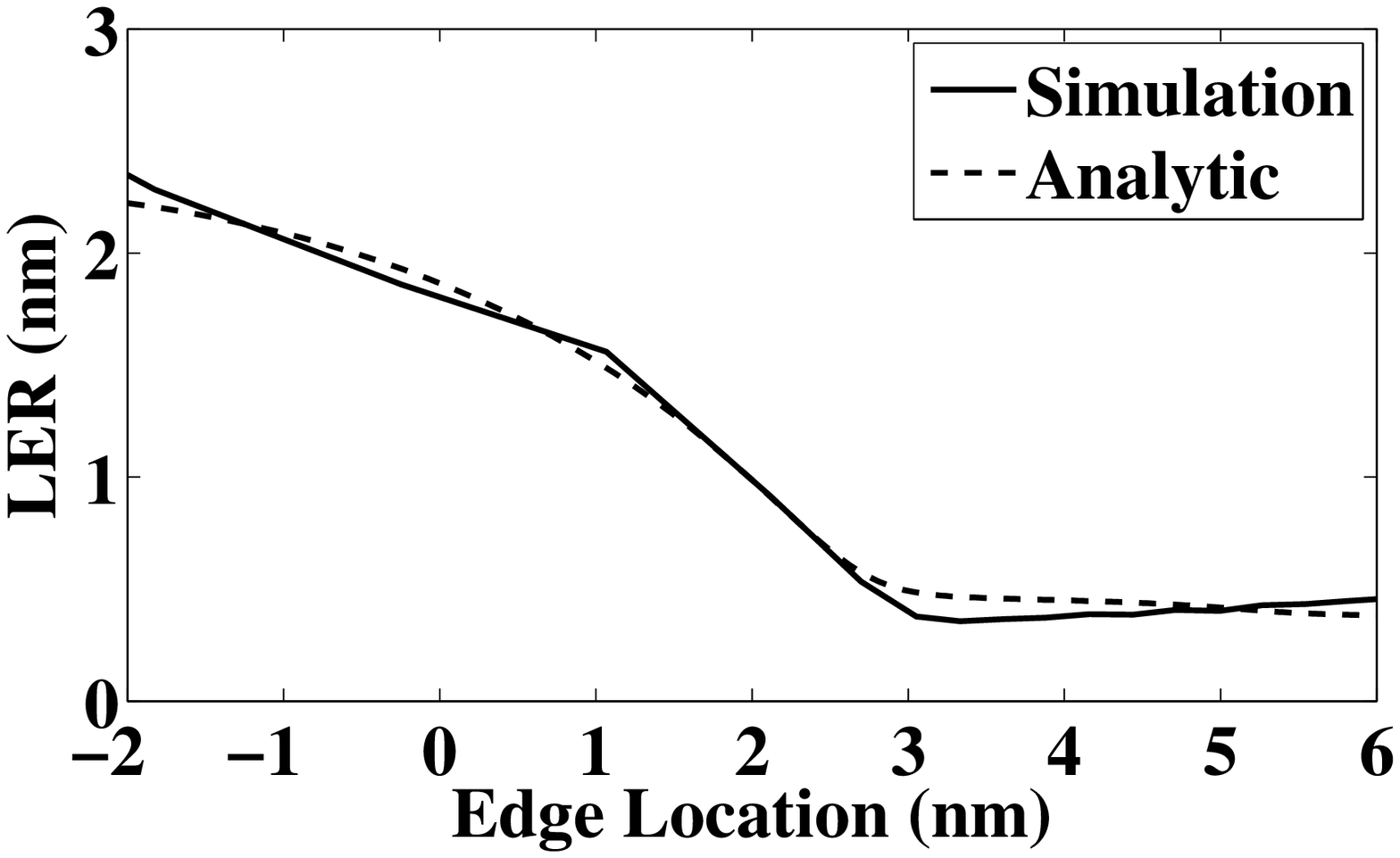} \\
(b)\\
\vspace{-0.2cm}
 \caption{LER results of large pattern estimated by the analytic model and simulation at the bottom layer at the corner: (a), and center: (b). Substrate system consists of 300nm PMMA on Si. The e-beam energy is 50keV and dose is 640 $\mu C/cm^2$, where pixel interval is 5nm and feature. PSF's are generated by 1K electrons.
}
\label{fig:muti_line}
\end{figure}
Fig. \ref{fig:40eLER} shows another case of the LER and compared with simulation results with the proper scaling. The results have some differences which may be introduced by fluctuation in the reality. Notice that results generated by PSF's using 40 electrons will have more chance of fluctuation than the results using PSF's generated by 1k electrons. Fig. \ref{fig:DiffSize} shows the LER generated using a long line feature with different width ($W$). $W$ varies from 30nm, 60nm to 90nm. As can be seen in the figure, they are all very close to each other which also show that LER is relatively independent of feature size.

Fig. \ref{fig:muti_line} shows the LER results of large pattern. 41-line pattern is considered and each line has 25 nm width and 4000 nm long. The space between lines is 25 nm. Resist is using 300nm PMMA on Si, the e-beam energy is 50kev and dose is 640 $\mu C/cm^2$ where pixel interval is 5 nm. Two locations in the pattern are compared: corner and center. As shown in the figure, the LER at the corner is slightly larger than that at the center. One reason is because in the center, more points make contribution to its exposure and statistically its exposure should be more stable (less fluctuation)than that at the corner.

\section{Summary}
\label{sec:Sum}

A computer simulation is often employed in many lithography research and development efforts. Usually an accurate simulation required to carry out Monte Carlo simulation at relatively large proportion of points and can be very time-consuming. In this paper, a new analytical model which can derive relatively accurate result and greatly cut down the computation required greatly in the simulation method. It also reveals certain relationship between PSF's and exposure fluctuation which is a major effect on the LER. Based on the statistics not the whole information of PSF's set, fluctuation of exposure can be expressed as a simple result. From the fluctuation of exposure, other information such as develop rate, path location and the final LER can be derived. After certain scaling which reflects the effect between different paths, the analytical results can be very accurate with the simulation result. Also it is verified by different cases.

\end{document}